\documentclass{ws-procs9x6}
\usepackage{amsfonts}
\usepackage{amsmath}
\usepackage{amssymb}
\usepackage{chapterbib}
\usepackage{dcolumn}
\usepackage{rotating}
\usepackage[percent]{overpic}
\usepackage{booktabs}
\usepackage{float}
\begin{document}
\title{Social distancing strategies against disease spreading}
\author{L. D. Valdez$^\dag$, C. Buono, P. A. Macri and
  L. A. Braunstein \footnote{Also at Center for Polymer Studies,
    Boston University, CPS, 590 Commonwealth Av, Boston, Massachusetts
    02215, USA}} \address{Instituto de Investigaciones F\'isicas de
  Mar del Plata (IFIMAR)-Departamento de F\'isica, Facultad de
  Ciencias Exactas y Naturales, Universidad Nacional de Mar del
  Plata-CONICET, Funes 3350, (7600) Mar del Plata,
  Argentina.\\ \email{$^\dag\!$E-mail: ldvaldes@mdp.edu.ar}}

\begin{abstract}
The recurrent infectious diseases and their increasing impact on the
society has promoted the study of strategies to slow down the epidemic
spreading. In this review we outline the applications of percolation
theory to describe strategies against epidemic spreading on complex
networks. We give a general outlook of the relation between link
percolation and the susceptible-infected-recovered model, and
introduce the node void percolation process to describe the dilution
of the network composed by healthy individual, $i.e$, the network that
sustain the functionality of a society. Then, we survey two
strategies: the quenched disorder strategy where an heterogeneous
distribution of contact intensities is induced in society, and the
intermittent social distancing strategy where health individuals are
persuaded to avoid contact with their neighbors for intermittent periods
of time. Using percolation tools, we show that both strategies may
halt the epidemic spreading. Finally, we discuss the role of the
transmissibility, $i.e$, the effective probability to transmit a
disease, on the performance of the strategies to slow down the
epidemic spreading.
\end{abstract}

\keywords{Epidemics, Percolation, Complex Networks}
\bodymatter

\section{Introduction}

Increasing incidence of infectious diseases such as the SARS and the
recent A(H1N1) pandemic influenza, has led to the scientific community
to build models in order to understand the epidemic spreading and to
develop efficient strategies to protect the
society~\cite{Baj_01,Col_01,Lip_01,ah1n1}.  Since one of the goals of
the health authorities is to minimize the economic impact of the
health policies, many theoretical studies are oriented to establish
how the strategies maintain the functionality of a society at the
least economic cost.

The simplest model that mimics diseases where individuals acquire
permanent immunity, such as the influenza, is the pioneer
susceptible-infected-recovered (SIR)
model~\cite{Boc_01,And_01,Gra_01,New_05}. In this epidemiological
model the individuals can be in one of the three states: i)
susceptible, which corresponds to a healthy individual who has no
immunity, ii) infected, $i.e.$ a non-healthy individual and iii)
recovered, that corresponds to an individual who cannot propagate
anymore the disease because he is immune or dead. In this model the
infected individuals transmit the disease to the susceptible ones, and
recover after a certain time since they were infected.  The process
stops when the disease reaches the steady state, $i.e.$ , when all
infected individuals recover. It is known that, in this process, the
final fraction of recovered individuals is the order parameter of a
second order phase transition. The phase transition is governed by a
control parameter which is the effective probability of infection or
transmissibility $T$ of the disease.  Above a critical threshold
$T=T_c$, the disease becomes an epidemic, while for $T<T_c$ the
disease reaches only a small fraction of the population
(outbreaks)~\cite{New_05,Mil_02,Ken_01,Lag_02}.  The first SIR model,
called random mixing model, assumes that all contacts are possible,
thus the infection can spread through all of them. However, in
realistic epidemic processes individuals have contact only with a
limited set of neighbors. As a consequence, in the last two decades
the study of epidemic spreading has incorporated a contact network
framework, in which nodes are the individuals and the links represent
the interactions between them. This approach has been very successful
not only in an epidemiological context but also in economy, sociology
and informatics~\cite{Boc_01}.  It is well known that the topology of
the network, $i.e.$ the diverse patterns of connections between
individuals plays an important role in many processes such as in
epidemic spreading~\cite{Vol_02,Sal_01,Bog_01,Val_00}. In particular,
the degree distribution $P(k)$ that indicates the fraction of nodes
with $k$ links (or degree $k$) is the most used characterization of
the network topology.  According to their degree distribution,
networks are classified in i) homogeneous, where node's connectivities
are around the average degree $\langle k \rangle$ and ii)
heterogeneous, in which there are many nodes with small connectivities
but also some nodes, called hubs or super-spreaders, with a huge
amount of connections.  The most popular homogeneous networks is the
Erd\"os R\'enyi (ER) network~\cite{Erd_02} characterized by a Poisson
degree distribution $P(k)=e^{-\langle k\rangle}\langle k \rangle
^{k}/k!$.  On the other hand, very heterogeneous networks are
represented by scale-free (SF) distributions with $P(k) \sim k^{-
  \lambda}$, with $k_{min} < k < k_{max}$, where $\lambda$ represents
the heterogeneity of the network.  Historically, processes on top of
complex networks were focused on homogeneous networks since they are
analytically tractable. However, different researches showed that real
social~\cite{New_08,Bar_02}, technological~\cite{Bar_01,Fal_01},
biological~\cite{Egu_01,Jeo_01} networks, etc, are very heterogeneous.
%In particular, for the SIR model in heterogeneous networks $T_c$
%vanishes, which implies that social networks, that have a very
%heterogeneous degree distribution are very vulnerable to develop an
%epidemic.

Other works showed that the SIR model, at its steady state, is related
to link percolation~\cite{Gra_01,New_05,Mey_01,Ken_01}. In percolation
processes~\cite{Sta_01}, links are occupied with probability
$p$. Above a critical threshold $p=p_c$, a giant component (GC)
emerges, which size is of the order of the system size $N$; while
below $p_c$ there are only finite clusters.  The relative size of the
GC, $P_{\infty}(p)$, is the order parameter of a geometric second
order phase transition at the critical threshold $p_c$.  Using a
generating function formalism~\cite{Wil_01,Dun_01,New_03}, it was
shown that the SIR model in its steady state and link percolation
belong to the same universality class and that the order parameter of
the SIR model can be exactly mapped with the order parameter
$P_{\infty}(p=T)$ of link percolation~\cite{New_05}. For homogeneous
networks the exponents of the transitions have mean field (MF) value,
although for very heterogeneous network the exponents depend on
$\lambda$.

Almost all the researches on epidemics were concentrated in studying
the behavior of the infected individuals. However, an important issue
is how the susceptible network behaves when a disease
spreads. Recently, Valdez {\it et. al.}~\cite{Val_02,Val_01}
studied the behavior of the giant susceptible component (GSC) that is
the functional network, since the GSC is the one that supports the
economy of a society.  They found that the susceptible network also
overcomes a second order phase transition where the dilution of the
GSC during the first epidemic spreading can be described as a ``node void
percolation'' process, which belongs to the same universality class
that intentional attack process with MF exponents.

Understanding the behavior of the susceptible individuals allows to
find strategies to slow down the epidemic spread, protecting the
healthy network. Various strategies has been proposed to halt the
epidemic spreading. For example, vaccination programs are very
efficient in providing immunity to individuals, decreasing the final
number of infected people~\cite{Fer_01,Ban_01}. However, these
strategies are usually very expensive and vaccines against new strains
are not always available during the epidemic spreading. As a
consequence, non-pharmaceutical interventions are needed to protect
the society. One of the most effective and studied strategies to halt
an epidemic is quarantine~\cite{Lag_01} but it has the disadvantage
that full isolation has a negative impact on the economy of a region
and is difficult to implement in a large population. Therefore, other
measures, such as social distancing strategies can be implemented in
order to reduce the average contact time between individuals. These
``social distancing strategies'' that reduce the average contact
time, usually include closing schools, cough etiquette, travel
restrictions, etc. These measures may not prevent a pandemic, but
could delay its spread.

In this review, we revisit two social distancing strategies named,
``social distancing induced by quenched disorder''~\cite{Buo_01} and
``intermittent social distancing'' (ISD) strategy~\cite{Val_01}, which
model the behavior of individuals who preserve their contacts during
the disease spreading. In the former, links are static but health
authorities induce a disorder on the links by recommending people to
decrease the duration of their contacts to control the epidemic
spreading.  In the latter, we consider intermittent connections where
the susceptible individuals, using local information, break the links
with their infected neighbors with probability $\sigma$ during an
interval $t_b$ after which they reestablish the connections with their
previous contacts. We apply these strategies to the SIR model and
found that both models still maps with link percolation and that they
may halt the epidemic spreading. Finally, we show that the
transmissibility does not govern the temporal evolution of the
epidemic spreading, it still contains information about the velocity
of the spreading.

\section{The SIR model and Link Percolation}
\label{models}

One of the most studied version of the SIR model is the time
continuous Kermack-McKendrick~\cite{Ker_01} formulation, where an
infected individual transmits the disease to a susceptible neighbor at
a rate $\beta$ and recovers at a rate $\gamma$. While this SIR version
has been widely studied in the epidemiology literature, it has the
drawback to allow some individuals to recover almost instantly after
being infected, which is a highly unrealistic situation since any
disease has a characteristic recovering average time. In order to
overcome this shortcoming, many studies use the discrete Reed-Frost
model~\cite{Bai_01}, where an infected individual transmits the
disease to a susceptible neighbor with probability $\beta$ and
recovers $t_r$ time units after he was infected.  In this model, the
transmissibility $T$ that represents the overall probability at which
an individual infects one susceptible neighbor before recover, is
given by
\begin{equation}
T=\sum_{u=1}^{t_R}\beta(1-\beta)^{u-1}=1-(1-\beta)^{t_R}.
\label{Trans}
\end{equation}

It is known that the order parameter $M_I(T)$, which is the final
fraction of recovered individuals, overcomes a second order phase
transition at a critical threshold $T\equiv T_c$, which depends on the
network structure.

One of the most important features of the Reed-Frost model (that we
will hereon call SIR model) is that it can be mapped into a link
percolation process~\cite{Gra_01,New_05,Mil_01,Mey_01}, which means
that is possible to study an epidemiological model using statistical
physic tools. Heuristically, the relation between SIR and link
percolation holds because the effective probability $T$ that a link is
traversed by the disease, is equivalent in a link percolation process
to the occupancy probability $p$. As a consequence, both process have
the same threshold and belong to the same universality
class. Moreover, each realization of the SIR model corresponds to a
single cluster of link percolation.  This feature is particularly
relevant for the mapping between the order parameters $P_{\infty}(p=T)$
of link percolation and $M_{I}(T)$ for epidemics, as we will explain
below.

For the simulations, in the initial stage all the individuals are in
the susceptible state.  We choose a node at random from the network
and infect it (patient zero). Then, the spreading process goes as
follows: after all infected individuals try to infect their
susceptible neighbor with a probability $\beta$, and those individuals
that has been infected for $t_r$ time steps recover, the time $t$
increases in one. The spreading process ends when the last infected
individual recovers (steady state).

\begin{figure}[ht]
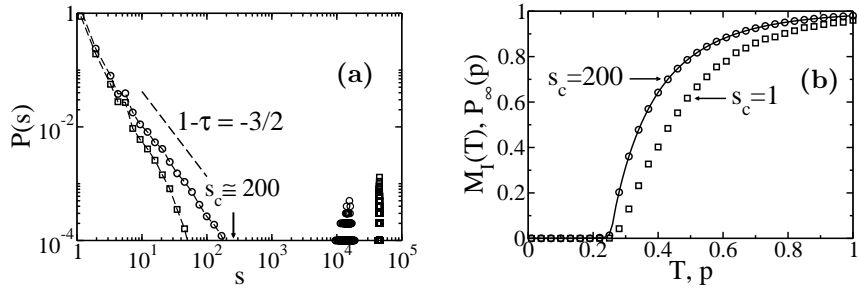

\centering
 \begin{overpic}[scale=0.20]{vbmbFig01.eps}
    \put(80,50){{\bf{(a)}}}
  \end{overpic}\hspace{0.5cm}
  \begin{overpic}[scale=0.20]{vbmbFig02.eps}
    \put(85,50){\bf{(b)}}
  \end{overpic}\vspace{0.5cm}
\caption{Effects of the cutoff $s_c$ on the mapping between the SIR
  model and link percolation for an ER network with $\langle k
  \rangle=4$ ($T_{c}=0.25$), $N=10^5$ . In (a) we show the probability
  $P(s)$ of a cluster of size $s$ (including the size of the giant
  component) in the SIR model for $T=0.27$ ($\bigcirc$) and $T=0.40$
  ($\square$). We can see that the gap between the epidemic sizes and
  the distribution of outbreaks increases with $T$. In Fig. (b) we show
  the simulation results for $M_{I}(T)$ for $s_c=1$ ($\square$) and
  $s_c=200$ ($\bigcirc$). Note that when $s_c=200$, we average the
  final size of infected clusters only over epidemic
  realizations. Considering only the conditional averages, we can see
  that $M_{I}(T)$ maps with $P_{\infty}(p)$ (solid line). Our
  simulations were averaged over $10^4$ realizations.}
\label{sc11}
\end{figure}

In a SIR realization, only one infected cluster emerges for any value
of $T$.  In contrast, in a percolation process, for $p < 1$ many
clusters with a cluster size distribution are generated~\cite{Mey_02}.
Therefore we must use a criteria to distinguish between epidemics (GC
in percolation) and outbreaks (finite clusters). The cluster size
distribution over many realizations of the SIR process, close but
above criticality, has a gap between small clusters (outbreaks) and
big clusters (epidemics).  Thus, defining a cutoff $s_c$ in the
cluster size as the minimum value before the gap interval, all the
diseases below $s_c$ are considered as outbreaks and the rest as
epidemics (see Fig.~\ref{sc11}a).  Note that $s_{c}$ will depend on
$N$.  Then, averaging only those SIR realizations whose size
exceeds the cutoff $s_c$, we found that the fraction of recovered
individuals $M_{I}(T)$ maps exactly with $P_{\infty}(p)$ (see
Fig.~\ref{sc11}b). For our simulations, we use $s_c=200$ for $N=10^5$.

It can be shown that using the appropriate cutoff, close to
criticality, all the exponents that characterizes the transition are
the same for both processes~\cite{Lag_02, Wu_01,Han_01}. Thus, above
but close to criticality
\begin{eqnarray}
M_I(T) &\sim& (T-T_c)^\beta,\\
P_{\infty}(p) &\sim& (p-p_c)^\beta,
\end{eqnarray}
with~\cite{Coh_02}
\begin{eqnarray}\label{eq.betat}
 \beta = \left\{
\begin{array}{ll}1 & \mbox{ for SF with $\lambda \geq 4$ and ER networks,}\\
\frac{1}{\lambda-3} & \mbox{ for $ 3 < \lambda < 4$,}
\end{array}
\right.
\end{eqnarray}
The exponent $\tau$ of the finite
cluster size distribution in percolation close to criticality is given by
 \begin{eqnarray}\label{eq.tau}
 \tau = \left\{
\begin{array}{ll} \frac{5}{2} & \mbox{ for SF with $\lambda \geq 4$ and ER networks;}\\
\frac{1}{\lambda-2}+2 & \mbox{ for $ 2 < \lambda < 4$.}
\end{array}
\right.
\end{eqnarray}
For the SIR model and for a branching process (see
Sec.~\ref{MathPerc}), there is only one ``epidemic'' cluster, thus
near criticality the probability of a cluster of size $s$, $P(s)$, has
exponent $\tau -1$, where $\tau$ is given by Eq.~(\ref{eq.tau}) (see
Fig.~\ref{sc11}a). For SF networks with $\lambda \leq 3$, in the
thermodynamic limit, the critical threshold is zero, and there is not
percolation phase transition.  On the other hand, for $\lambda \geq 4
$ and ER networks, all the exponents take the mean field (MF) values.

\section{Mathematical approach to link percolation}\label{MathPerc}

Given a network with a degree distribution $P(k)$, the probability to
reach a node with a degree $k$ by following a randomly chosen link on
the graph, is equal to $k P(k) /\langle k \rangle$, where $\langle k
\rangle$ is the average degree. This is because the probability of
reaching a given node by following a randomly chosen link is
proportional to the number of links $k$ of that node and $\langle k
\rangle$ is needed for normalization.  Note that, if we arrive to a
node with degree $k$ following a random chosen link , the total number
of outgoing links or branches of that node is $k - 1 $ .  Therefore,
the probability to arrive at a node with $k-1$ outgoing branches by
following a randomly chosen link is also $k P(k)/\langle k \rangle$. This
probability is called excess degree probability~\cite{New_07,Bra_01}.

In order to obtain the critical threshold of link percolation, let us
consider a randomly chosen and occupied link. We want to compute the
probability that through this link an infinite cluster {\it cannot} be
reached. For simplicity, we assume to have a Cayley tree. Here we will
denote a Cayley tree as a {\it single} tree with a given degree
distribution. Notice that link percolation can be thought as many
realizations of Cayley tree with occupancy probability $p$, which give
rise to many clusters. By simplicity we first consider a Cayley tree
as a deterministic graph with a fixed number $z$ of links per
node. Assuming that $z=3$, the probability that starting from an
occupied link we cannot reach the $n-th$ shell through a path composed
by occupied links, is given by
\begin{equation}
Q_n(p) =\left[(1-p) + pQ_{n-1}(p)\right]^2.
\end{equation}
Here, the exponent $2$ takes into account the number of outgoing links
or branches, and $1-p+p\; Q_{n-1}(p)$ is the probability that one
outgoing link is not occupied plus the probability that the link is
occupied ($i.e.$, at least one shell is reached) but it cannot lead to
the following $nth-1$ shell~\cite{Boc_01}. In the case of a Cayley
tree with a degree distribution, we must incorporate the excess degree
factor which accounts for the probability that the node under
consideration has $k-1$ outgoing links and sum up over all possible
values of $k$.  Therefore, the probability to {\it
  not reach} the generation $n-th$ can be obtained by applying a
recursion relation
\begin{eqnarray}
Q_n(p) &=& \sum_{k=1}^\infty \frac{k\;P(k)}{\langle k \rangle}
\left[(1-p)+pQ_{n-1}(p)\right]^{k-1},\\ &=& G_{1}[(1-p)+pQ_{n-1}(p)],
\end{eqnarray}
where $G_{1}(x)=\sum_{k=1}^{\infty}kP(k)/\langle k \rangle x^{k-1}$ is
the generating function of the excess degree distribution. As $n$
increases, $Q_{n}\approx Q_{n-1}$ and the probability that we cannot
reach an infinite cluster is
\begin{eqnarray}\label{Qinf}
Q_\infty(p) &=& G_{1}[(1-p)+pQ_{\infty}(p)].
\end{eqnarray}
Thus, the probability that the  starting link connects to an infinite
cluster is $f_{\infty}(p)=1-Q_{\infty}(p)$. From Eq~(\ref{Qinf}),
$f_{\infty}(p)$ is given by
\begin{eqnarray}\label{finfeq}
f_{\infty}(p)&=&1- G_{1}[1-pf_{\infty}(p)].
\end{eqnarray}

\begin{figure}[ht]
\centering
\includegraphics[scale=0.23]{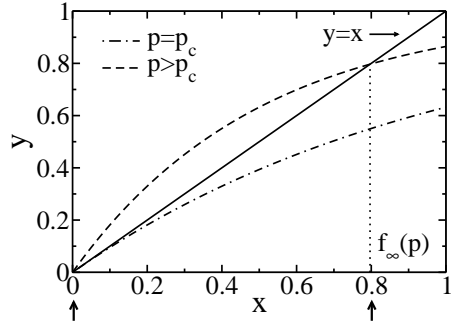}
\caption{ Geometrical solution of Eq. (\ref{finfeq}).  The straight
  line $y=x$ represents the left hand side of the equation. The
  dot-dashed line represents the right hand side (r.h.s) for $p=p_c$,
  where the r.h.s. is tangential to $y=x$ at the origin.  The
  dashed curve represents the r.h.s. for $p>p_c$. The vertical arrows
  indicate the points at which the identity function intersects with
  $y=1-G_{1}(1-px)$. Both cases are computed for the Poisson degree
  distribution with $\langle k \rangle=4$.
\label{f:iterations}}
\end{figure}
The solution of equation can be geometrically understood in
Fig.~\ref{f:iterations} as the intersection of the identity line $y=x$
and $y=1-G_{1}(1-px)$, which has at least one solution at the origin,
$x=f_\infty(p)=0$, for any value of $p$. But if the derivative of the
right hand side of Eq.~(\ref{finfeq}) with respect to $x$,
$\left[1-G_1(1-px)\right]'\vert_{x=0}=pG_1'(1)>1$,
we will have another solution in $0<x\leq 1$. This
solution $x=f_{\infty}(p)$ has the physical meaning of being the
probability that a randomly selected occupied link is connected to an
infinite cluster. The criticality corresponds to the value of
$p=p_{c}$ at which the curve $1-G_{1}(1-px)$ has exactly slope equal
one. Thus $p_c$ is given by~\cite{Coh_01}
\begin{equation}
p_c\equiv {1\over G_1'(1)}={\langle k\rangle\over \langle k^2\rangle -
  \langle k \rangle}.
\label{eq:pc}
\end{equation}
For ER networks, we have $p_{c}=1/\langle k \rangle$.  On the other
hand, we can obtain the order parameter of link percolation
$P_{\infty}(p)$, which represents the fraction of nodes that belongs
to the giant cluster when a fraction $p$ of links are occupied in a
random Cayley tree. The probability that a node
with degree $k$ does not belong to the giant component is given by the
probability that none of its links connect the node to the GC,
$i.e.$, $\left[1-pf_{\infty}(p)\right]^k$. Thus the fraction of nodes
that belong to the GC is
$1-\sum_{k=0}^{\infty}P(k)\left[1-pf_{\infty}(p)\right]^k$. Since the
relative epidemic sizes in the SIR model maps exactly with the
relative size of the giant component, we have that
\begin{equation}
M_{I}(T)=P_\infty(p=T)=1-G_0\left[1-p f_\infty(p)\right],
\end{equation}
where $G_{0}(x)=\sum_{k=0}^{\infty}P(k)x^{k}$ is the generating
function of the degree distribution and $ f_\infty(p)$ is the
non-trivial solution of Eq. (\ref{finfeq}) for $p>p_{c}$. It is
straightforward to show that for ER networks
$G_0(x)=G_1(x)=\exp{\left[-\langle k \rangle (1-x)\right]}$ and thus
$f_{\infty}(p)=P_{\infty}(p)$.  For pure SF networks, with $1\leq k <
\infty$, the generating function of the excess degree distribution is
proportional to the poly-logarithm function
$G_1(x)=Li_{\lambda}(x)/\xi(\lambda)$, where $\xi(\lambda)$ is the
Riemann function~\cite{Bra_01}.

In the current literature, the epidemic spreading is usually described
in terms of compartmental quantities, such as the fraction of infected
or susceptible individuals during an epidemic, and very little has
been done to describe how the disease affects the topology of the
susceptible network that can be considered as the functional network.
In the following section, we explain how an epidemic affects the
structure of the functional network in the steady state.

\section{Node Void Percolation and the SIR model}\label{Node_Void}
We define ``active'' links as those links pairing infected and
susceptible individuals.  During the epidemic spreading, the disease
is transmitted across active links, leading in the steady state to a
cluster composed by recovered individuals and clusters of susceptible
individuals. Alternatively, the growing process of the infected
cluster can also be described as a dilution process from the
susceptible point of view. Under this approach, as the ``infectious''
cluster grows from a root, the sizes of the void clusters, $i.e.$
those clusters composed by susceptible individuals, are reduced as in
a node dilution process, since when a link is traversed a void cluster
loses a node and all its edges. However, the susceptible nodes are not
randomly uniform reached by the disease because they are chosen
following a link. As a consequence higher degree nodes are more likely
to be reached than the ones with small degrees. We will call ``node
void percolation'' to this kind of percolation process in which the
void nodes are not removed at random. In this dilution process, there
exists a second critical value of the transmissibility $T^*$ (with
$T^*>T_c$), above which the giant susceptible component (GSC) is
destroyed.

Similarly to link percolation, in a Cayley tree (branching process)
the analytical treatment for the dilution of the susceptible network
uses a generating function formalism, that allows to compute the
existence of a GSC and its critical threshold.

Considering the same growing infected cluster process as in the
previous section, for large generations $f_{\infty}(p=T)$ can also be
interpreted as the probability that starting from a random chosen
link, a path or branch leads to the GC. Thus, if we cannot reach a GC
through a link, as we have a single tree, that link leads to a void
node. Thus the probability $V^s$ to reach a void node through a link
is given by
\begin{eqnarray}
V^{s}= 1-f_{\infty}(T)=G_1\left[1-p f_{\infty}(T)\right]
\end{eqnarray}
which is also the probability to reach a susceptible individual by
following a link at a given transmissibility $T$. It was shown that
$V^{s}$ is a fundamental observable to describe the temporal evolution
of an epidemic~\cite{Val_02,Mil_04,Mil_03}. As in the usual
percolation process, there is a critical threshold $V^{s}_c$ at which
the susceptible network undergoes a second order phase
transition. Above $V^{s}_c$ a GSC exists while at and below $V^{s}_c$
susceptible individuals belong only to finite components. As a
consequence, the transmissibility $T^{*}$ needed to reach this point
fulfills~\cite{Val_01}
\begin{equation}\label{condition_p}
V_{c}^{s}=G_1[1-T^{*} f_{\infty}(T^{*})].
\end{equation}
Therefore, from Eq~(\ref{condition_p}) we obtain
the self consistent equation
\begin{equation}\label{pestrellaFinal}
V_{c}^{s}=G_{1}\left[1-T^{*}(1-V_{c}^{s})\right],
\end{equation}
where $T^{*}$ is the solution of Eq.~(\ref{pestrellaFinal}) and
$V_{c}^{s}$ is given by
$V_{c}^{s}=G_{1}[(G_1^{'})^{-1}(1)]$~\cite{Val_02} as can be seen in
Appendix A and Ref.~\cite{Val_02}. Thus for a virulent disease with
$T\geq T^{*}>T_{c}$, we have $V^{s}<V^{s}_c$ and therefore the size of
the GSC $S_1\to0$~\cite{Val_01}. The theoretical value of $S_{1}$ for
a given value of $V^{s}$ can be obtained using an edge-based
compartmental approach~\cite{Mil_04,Mil_03,Val_02} that it is
explained in Appendix A.

\begin{figure}[ht]
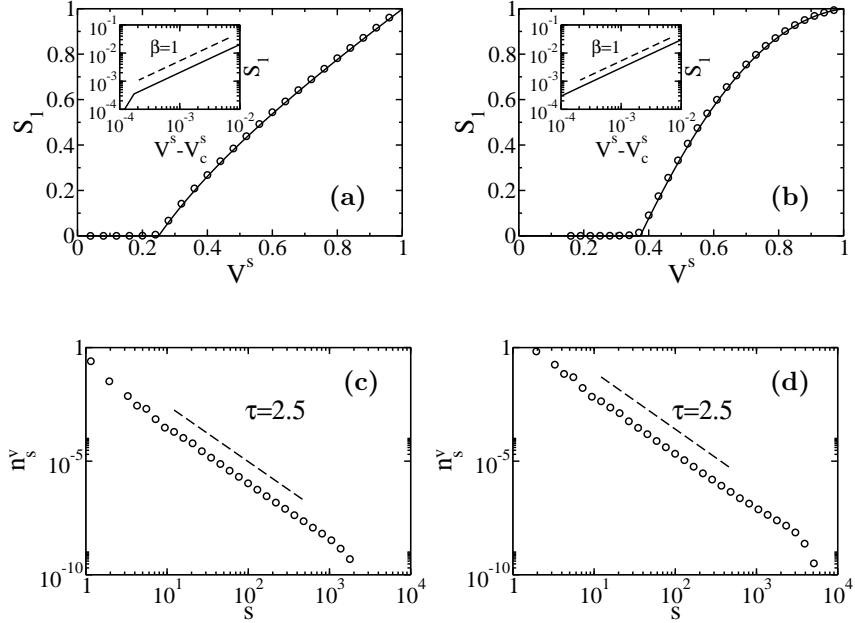

\centering
\vspace{1.0cm}
  \begin{overpic}[scale=0.20]{vbmbFig04.eps}
    \put(80,20){{\bf{(a)}}}
  \end{overpic}\hspace{0.5cm}
  \begin{overpic}[scale=0.20]{vbmbFig05.eps}
    \put(80,20){\bf{(b)}}
  \end{overpic}\vspace{0.7cm}
  \begin{overpic}[scale=0.20]{vbmbFig06.eps}
    \put(80,55){\bf{(c)}}
  \end{overpic}\vspace{0.5cm}
  \begin{overpic}[scale=0.20]{vbmbFig07.eps}
    \put(80,55){\bf{(d)}}
  \end{overpic}\vspace{0.5cm}
\caption{Fraction $S_1$ of nodes belonging to the GSC, as a function
  of $V^s$ for $N=10^5$ in an ER network and $\langle k \rangle=4$ (a)
  and SF network with $\lambda=2.63$, $k_{min}=2$ with $\langle k
  \rangle =4$ (b). The solid lines correspond to the solution of
  Eqs.~(\ref{eqs111}-\ref{eqs113}) and simulations are in symbols. In
  the insets, we show the power-law behavior of $S_1$ with the
  distance to the criticality $V^{s}_{c}$. Similarly, in figures (c)
  and (d) we plot the void node cluster size distribution at
  $V^{s}_{c}$ for ER ($V^{s}_{c}=1/4$) and SF networks
  ($V^{s}_{c}=0.38$), respectively. For homogeneous and heterogeneous
  networks the critical exponents are always those of MF [see
    Eq.(\ref{eq.betat}-\ref{eq.tau})] with values $\beta=1$ and
  $\tau=5/2$. }\label{S111}
\end{figure}
When $V^{s}\to V^{s}_{c}$, the size of the giant component $S_{1}$ and
the distribution of void cluster's sizes $n_s^{v}$, behave with the
distance to criticality as power laws.
\begin{eqnarray}
S_{1}&\sim& (V^{s}-V^{s}_{c})^{\beta}, \;\;\;\mbox{for $V^{s} \gtrsim V^{s}_{c}$,}\\
n_s^{v} &\sim& s^{\tau},\;\;\; \mbox{at $V^{s}_{c}$},
\end{eqnarray}
but in contrast to link percolation, their critical exponents have MF
values, $i.e.$, $\beta=1$ and $\tau=5/2$ for homogeneous and
heterogeneous networks [see Fig.~\ref{S111} and
Eqs.~(\ref{eq.betat}-\ref{eq.tau})]. Since two critical exponents are
enough to characterize a phase transition, then all the critical
exponents have MF values, as in an intentional attack percolation
process independently of the network's topology~\cite{Coh_04,Val_02}.

These results are not only restricted to the steady state, but also
can be extended to the temporal evolution of an epidemic spreading. It
can be shown that during the spreading, the GSC dilutes as in a node
void percolation process. In particular, for $T>T^{*}$, there exists a
critical time at which the GSC has the second order transition that we
explained before. For further details, see Ref.~\cite{Val_02}

All the concepts and tools previously introduced provide the basis for
the study of the spread of an epidemic and the evolution of the GSC,
that will be applied to the analysis of strategies against the
epidemic spreading.

\section{Social distancing induced by quenched disorder}

Living in society implies that individuals are constantly interacting
with each other. Interactions may take different forms, but those
involving proximity or direct contact are of special interest because
they are potential bridges to propagate infections. Empirical data
suggest that human contacts follow a broad distribution
\cite{Kar_01,Catt_01,Catt_02}. These results support the idea that
social interactions are heterogeneous, that means that individuals
have a lot of acquaintances but just a few of them are close
contacts. This heterogeneity between contacts can be thought as a
network with quenched disorder on the links, wherein the disorder is
given by a broad distribution. For example, if the weights represent
the duration of the contacts between two
individuals~\cite{Bra_01,ops,barrat}, the larger the weight, the
easier is for an infection to traverse the link.

An important feature of the networks topology without disorder is the
shortest average distance $\ell$, defined as the minimum average
number of connections between all pairs of nodes, which behaves as
$\ell\sim \ln(N)$ for ER networks~\cite{Strog_01} and as $\ln\ln(N)$
for very heterogeneous networks. This is why these networks are the
called small or ultra small world~\cite{Coh_05}. It is known that the
disorder can dramatically alter some topological properties of
networks. Several studies have shown that when the disorder between
connections is very broad or heterogeneous, also called strong
disorder limit (SD), the network loses the small world property and
the average distance goes as a power of $N$ for ER and SF networks
with $\lambda>3$ due to the fact that the SD can be related to
percolation at criticality \cite{bra01,bra03,scre,porto}. However, the
exact mapping between the order parameter of both second order phase
transitions of percolation and SIR is not affected by a random
disorder.

In the real life, the disorder in the network can be modified by
health policies in order to, for example, delay the disease spreading
allowing the health services to make earlier interventions
\cite{Buo_01}. Using different methods like broadcasting, brochures or
masks distributions, the public health agencies can induce people to
change their effective contact time and therefore the heterogeneity of
the interactions. This strategy was tacitly used by some governments
in the recent wave of influenza A(H1N1) epidemic in 2009 \cite{ah1n1},
but until now the effectiveness of the strategy and how it depends on
the virulence and the structure of the disease has not been widely
studied.

We study how the heterogeneity of the disorder affects the disease
spreading in the SIR model for a theoretical quenched disorder
distribution with a control parameter for its broadness. Using a
theoretical disorder distribution given by,
\begin{equation}
P(w)=\frac{1}{aw} \;,
\label{eqPw}
\end{equation}
where $P(w)\neq 0$ in [$e^{-a}$,1], and $a$ is the parameter which
controls the width of the weight distribution and determines the
strength of the disorder. Note that as $a$ increases, more values of
the weight are allowed and thus the distribution is more
heterogeneous.

In our weighted model the spreading dynamics follow the rules of the
SIR model explained in Sec.~\ref{models}, with a probability of
infection that depends on the weight of each link, such that each
contact in the network has infection probability $\beta w$, where
$\beta$ represents the virulence characteristic of the disease in
absence of disorder.

This type of weight has been widely used
\cite{bra03,Cie_01,porto,bra02} and it is a well known example of
many distributions that allow to reach the strong disorder limit in
order to obtain the mapping with percolation. With this weight
distribution the transmissibility $T(\beta,t_r,a)=T_{a}$ is given by
Eq.~(\ref{Trans}) replacing $\beta$ by $\beta w$ and integrating over
the weight distribution~\cite{San_01}, thus
\begin{eqnarray}
T_{a}&=&\sum_{u=1}^{t_r}\int_{e^{-a}}^{1}\beta w \frac{(1-w \beta)^{u-1}}{aw}dw \nonumber\\
&=& \sum_{u=1}^{t_r}\frac{\left(1-\beta e^{-a}\right)^{u}-\left(1-\beta\right)^{u}}{a\;u},
\label{Tdisorder}
\end{eqnarray}
Note that, in the limit of $a \to 0$ we recover the classical SIR
model (non disordered) with a fixed infection probability $\beta$ with
$T=1-(1-\beta)^{t_r}$. When $a\to \infty$ there will be links in the
network with zero weight and the strategy turns to a total quarantine
with $T_{a}\to 0$. For example, if $t_r=1$, $T_{a} =\beta (1-e^{-a})/a
\simeq \beta/a$ with $a >> 1$, thus the transmissibility $T_{a}$ will
be smaller than the intrinsic transmissibility $T$ of the disease
without strategy for any $a > 0$, reducing the epidemic spreading.

\begin{figure}
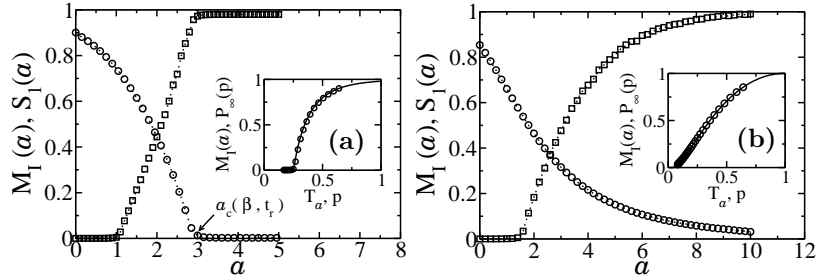

\centering
\vspace{1.0cm}
  \begin{overpic}[scale=0.20]{vbmbFig08.eps}
    \put(80,32){{\bf{(a)}}}
  \end{overpic}\vspace{0.5cm}
  \begin{overpic}[scale=0.20]{vbmbFig09.eps}
    \put(80,32){\bf{(b)}}
  \end{overpic}\vspace{0.5cm}
\caption{Linear-linear plots of the mass of recovered individuals
  $M_I(a)$ ($\circ$) and $S_{1}(a)$ ($\square$) in the steady state of
  the epidemic spreading as a function of the strength parameter of
  the disorder $a$ for $N=10^{5}$, $\beta=0.05$ and $t_r=20$ in an ER
  network with $\langle k\rangle=4$ (a) and SF network with
  $\lambda=2.63$ (b). Dotted lines are given as guides for the
  eye. Note that without disorder, the transmissibility is $T\simeq
  0.64$, and as $a$ increases the effective transmissibility $T_{a}$
  decreases, and the disease gets less virulent. The insets shows
  $M_I(a)$ from the main plot and $P_\infty$ as a function of $T_a$
  and $p$ showing the exact mapping between our model and
  percolation. Our simulations were averaged over $10^4$
  realizations.}
 \label{Fig-dis1}
\end{figure}

In the following, we only consider those propagations that lead to
epidemic states, and disregard the outbreaks. As the substrate for the
disease spreading we use both, ER and SF networks. After the system
reaches the steady state, we compute the mass of recovered individuals
$M_I(a)$ and the size of the functional network $S_{1}(a)$ as a
function of $a$. Given an intrinsic transmissibility $T$ of the
disease before the strategy is applied (see Eq.~(\ref{Trans})), as $a$
increases, the impact of the disease on the population decreases as
shown in Fig~\ref{Fig-dis1}. We can see that in ER networks
Fig~\ref{Fig-dis1}(a) there is a threshold $a = a_c(\beta,t_r)$ above
which the epidemic can be stopped and only outbreaks occurs (epidemic
free phase). However for very heterogeneous SF networks
Fig~\ref{Fig-dis1}(b), $a_c(\beta,t_r)$ must increase noticeably in
order to stop the epidemic spreading. For the steady
magnitudes, the SIR process is always governed by the effective
transmissibility $T_a$ given by Eq.~(\ref{Tdisorder}), as shown in the
inset of Fig.~\ref{Fig-dis1}.

\begin{figure}
 \centering
\vspace{0.5cm}
 \includegraphics[scale=0.23]{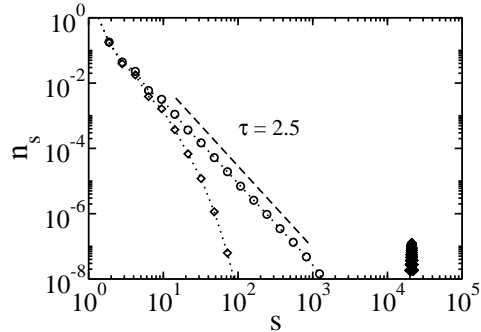}
 \caption{Cluster size distribution of the susceptible individuals for
   $\beta=0.05$ and $t_r=20$. Circles correspond $T_{a}=0.46$ with
   $a=1.0$ where there are clusters of all sizes of susceptible
   individuals. The dashed line is a fitting from which $n_s\sim s^{-2.5}$
   and is set as a guide to the eye. The diamonds correspond to
   $T_{a}=0.40$ with $a=1.5$ for which susceptible individuals show a
   herd behavior. Our simulations were averaged over $10^4$
   realizations.}
 \label{Fig-dis2}
\end{figure}

With the disorder strategy, the contact time between infected and
susceptible individuals decreases hindering the disease spreading and
protecting the functional network. We will refer to this defense
mechanism of healthy individuals as ``susceptible herd behavior''. As
explained in Sec.~\ref{Node_Void}, there is a $T^*$ that is the
solution of Eq.~(\ref{pestrellaFinal}) below which the susceptible
herd behavior generates a GSC. In Fig.~\ref{Fig-dis2} we show the
cluster size distribution of the susceptible individuals $n_s$ for
$T_{a}\simeq T^*$ and for $T_{a}<T^*$ for ER networks, which show that
the exponent $\tau=5/2$ takes the mean field value of node
percolation.

\begin{figure}
 \centering
\vspace{0.5cm}
\includegraphics[scale=0.25]{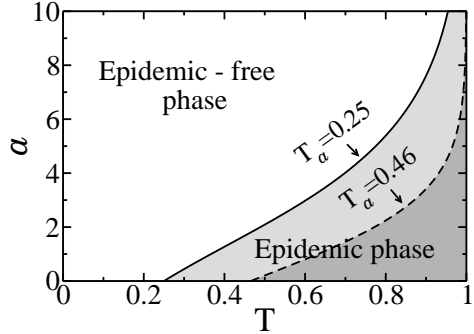}
 \caption{Plane $T-a$ for the SIR model with $t_r=20$ and infection
   probability distribution for each contact $\beta w$ with weight
   distribution $P(w)=1/a w$ in $[e^{-a},1]$. The solid line that
   corresponds to $T_a =1/4$ that is $T_c$ for an ER network with
   $\langle k \rangle =4$, separates the epidemic phase from the
   epidemic free phase region shown in dark gray. The dashed line
   shows $T_a = 0.46$ that is $T^*$ below which a giant component of
   susceptible emerges. The light gray region is the phase in which
   the GSC and the giant recovered cluster coexists.}
 \label{Fig-dis3}
\end{figure}

In Fig.~\ref{Fig-dis3} we plot the plane $T-a$ in order to show how
$T_a$ depends on the intrinsic transmissibility of the disease $T$ and
on the heterogeneity of the disorder $a$. The full line in the plane
$T-a$ corresponds to a $T_a=T_c=0.25$, and separates the epidemic free
phase (non colored region) from the epidemic phase (dark gray
region). Note that $a$ is a parameter that could be controlled by the
authorities, therefore the plane $T-a$ shows the required
heterogeneity of the disorder needed to avoid an epidemic spreading
depending on the virulence of the disease, characterized by the
intrinsic $T$. The dashed line corresponds to a $T_a=T^*$, below which
a GSC emerges. The light gray area indicates the phase where there is a
coexistence of giant clusters of infected and susceptible individuals.

In this strategy, there are no restrictions on which individual to get
away from. Another strategy could be to advise people to cut
completely their connection with their infected contacts (when
possible) for a given period of time. This kind of strategy will be
analyzed in the next section.

\section{Intermittent Social Distancing Strategy}\label{Sec_Inf_Cur}

In the previous strategy, individuals set a quenched disorder on
the intensity of the interaction with their neighbors in order to
protect themselves from the epidemic spreading. An alternative
strategy consists of susceptible individuals that inactivate the
interactions with their infected neighbors, but reestablish their
contacts after some fixed time. This strategy that we call
intermittent social distancing (ISD) strategy mimics a behavioral
adaptation of the society to avoid contacts with infected individuals
for a time interval, but without losing them permanently. This is an
example of adaptive network where the topology coevolves with the
dynamical process~\cite{Gro_01, Gro_02}.

Specifically, we study an intermittent social distancing strategy
(ISD) in which susceptible individuals, in order to decrease the
probability of infection, break (or inactivate) with probability
$\sigma$ their links with infected neighbors for intermittent periods
of length $t_b$.

We closely follow the presentation of this model from
Ref.~\cite{Val_01}.  Assuming that the disease spreads with
probability $\beta$ through the active links and that the infected
individuals recovers after $t_r$ time steps, at each time step the
infected individual tries first to transmit the disease to his
susceptible neighbors, and then if he fails, susceptible individuals
break their links with probability $\sigma$ for a period $t_b$.

These dynamic rules generate an intermittent connectivity between
susceptible and infected individuals that may halt the disease
spreading. In the limit case of $t_b>t_r$, the ISD strategy is
equivalent to a permanent disconnection, because when the link is
restored the infected neighbor is recovered (or dead) and cannot
transmit the disease anymore.

In order to compute the transmissibility for this strategy, we first
introduce the case $\sigma=1$ and then we generalize for any value of
$\sigma$. For the case $\sigma=1$, let consider that an active link
appears and denote the first time step of its existence as $m=1$. At
this time step, the active link tries to transmit the disease with
probability $\beta$, if it fails that link will be broken for the next
$t_b$ time steps. After restoring that active link, the process is
periodically repeated with period $t_b+1$, until the disease is
transmitted or the infected individual recovers. On the other hand,
the time steps at which the link is active are located at times
$m=(t_b+1)u+1$ where $u$ is an integer number defined in the interval
$0\leq u \leq [(t_r-1)/(t_b+1)]$, where $u=0$ corresponds to the first
time step, and $[(t_r-1)/(t_b+1)]$ is the maximum number of
disconnection periods that leaves at the end at least one time step to
transmit the disease. In particular, the probability to transmit the
disease at the next time after $u$ disconnection periods is given by
$\beta (1-\beta)^u$. Then summing over all possible values of $u$, the
total transmissibility $T(\beta,\sigma,t_r,t_b)\equiv
T_{\sigma}$~\cite{Val_01} is given by
\begin{eqnarray}\label{sigma1}
T_{\sigma}&=&\beta \left(1+\sum_{u=1}^{\left[\frac{tr-1}{t_{b}+1}\right]}(1-\beta)^{u}\right),\nonumber\\
&=&1-\left(1-\beta\right)^{\left[\frac{t_r-1}{t_b+1}\right]+1}.
\end{eqnarray}

For the case $0<\sigma<1$, first consider the example with only one
disconnection period ($u=1$), $t_r=10$, $t_b=2$ and the infectious
transmission at the time step $m=8$, that is illustrated in the first
line of Table~\ref{table1}. Note that in this case, there are only
$m-u\;t_b=6$ time units at which the link is active. Then, for this
example the transmissibility is proportional to four factors: i)
$\beta(1-\beta)^{5}$ since there are $5$ active time steps at which
the infected individual cannot transmit the disease, and at the last
time unit the disease is transmitted, ii) $\sigma$, because the link
is broken one time, iii) $(1-\sigma)^{4}$, because during $6$ active
time steps the infected individual does not break the link except just
before each inactive period and the last day, and iv)
$\binom{m-u\;t_{b}-1}{u}=\binom{5}{1}=5$ that is the total number of
configurations in which we can arrange one inactive period in a period
of length $7$ (this factor only takes into account the first $m-1=7$
time units, because the disease is transmitted at time $m=8$. See the
first line of Table~\ref{table1}).
\begin{table}
\tbl{Disconnected periods for a pair $S-I$ with $t_r=10$ (recovery
  time), $t_b=2$ (disconnection period) and $m=8$ (time of
  infection). The first column represents the number of disconnected
  periods $u$ before $m=8$, the second column is a typical
  configuration, the third column is the probability of that
  configuration and the fourth column is the number of ways to arrange
  $u$ disconnected periods. In the second column, each cell correspond
  to a time unit. The white cells represent the time units where a
  link between the $S$ and the $I$ node exists, the gray ones
  correspond to the disconnection period and in the black cells there
  is no dynamic for the pair $S-I$ because the $S$ has been infected
  and now the pair becomes $I-I$. Notice that initially the link
  cannot be broken because this disconnection only happens after that
  the $I$ individual fails to infect the susceptible one, with
  probability $(1-\beta)$. Similarly, two disconnection periods must
  be separated by at least one white cell. \label{table1}}
{%\renewcommand{\arraystretch}{1}
\begin{tabular}{cccc}  %% el @{} es para agregar espacio
\toprule[0.03cm]
\multicolumn{1}{c}{u}&\multicolumn{1}{c}{Example}&\multicolumn{1}{c}{Probability}&\multicolumn{1}{c}{Binomial}\\
\multicolumn{1}{c}{}&\multicolumn{1}{c}{}&\multicolumn{1}{c}{}&\multicolumn{1}{c}{Coefficient}\\
\cmidrule [0.03cm]{1-4}
\multicolumn{1}{c}{$u=1$}&\multicolumn{1}{m{0.5cm}} {\includegraphics[scale=0.50]{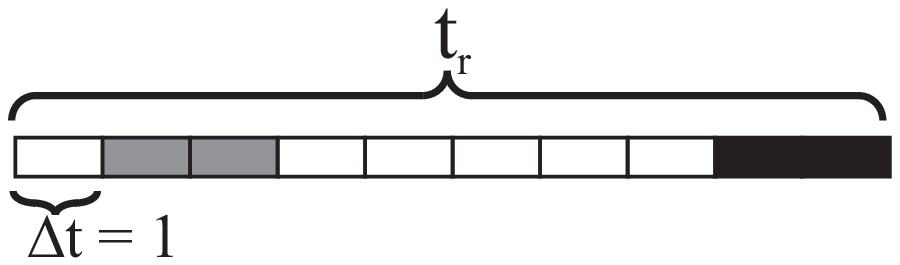}}&\multicolumn{1}{c}{$\beta\;\sigma(1-\sigma)^{4}(1-\beta)^{5}$}&$\binom{8-2-1}{1}=5$\\\cmidrule{1-4}
\multicolumn{1}{c}{$u=2$}&\multicolumn{1}{c}{\includegraphics[scale=0.50]{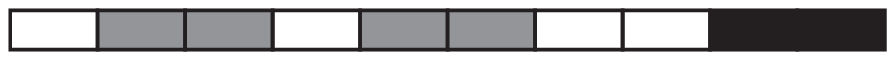}}&\multicolumn{1}{c}{$\beta\;\sigma^{2}(1-\sigma)^{1}(1-\beta)^{3}$}&$\binom{8-4-1}{2}=3$   \\
\bottomrule[0.03cm]
\end{tabular}}
\end{table}
%m{0.5cm}
In the general case, for all the values $0 < m \leq tr$, the disease
spreads with a total transmissibility given by,
\begin{eqnarray}\label{Ec.Trans}
T_{\sigma}=\sum_{m=1}^{t_r} \beta(1-\beta)^{m-1}(1-\sigma)^{m-1}+\beta\sum_{m=t_{b}+2}^{t_r}\phi(m,t_b,\sigma,\beta).
\end{eqnarray}
In the first term of Eq.~(\ref{Ec.Trans}),
$\beta(1-\beta)^{m-1}(1-\sigma)^{m-1}$ is the probability that an
active link is lost due to the infection of the susceptible individual
at time step $m$ given that the active link has never been broken in
the $m-1$ steps since it appears. In the second term of
Eq.~(\ref{Ec.Trans}), $\beta\;\phi(m,t_b,\sigma,\beta)$ denotes the
probability that an active link is lost due to the infection of the
susceptible individual at time $m$ given that the link was broken at
least once in the first $m-1$ time units. The probability
$\phi(m,t_b,\sigma,\beta)$, which is only valid for $m\geq t_b+2$ is
given by~\cite{Val_01}
\begin{eqnarray}\label{eqSumPhi}
\phi(m,t_b,\sigma,\beta)\equiv\phi_{m}&=&\sum_{u=1}^{\left[\frac{m-1}{t_{b}+1}\right]}\binom{m-u\;t_{b}-1}{u}\sigma^{u}\times \nonumber\\
&&(1-\sigma)^{m-1-u(t_{b}+1)}(1-\beta)^{m-1-u\;t_{b}},
\end{eqnarray}
where $\left[\;\cdot \;\right]$ denotes the integer part function.

With the ISD strategy~\cite{Val_01} the effective probability of
infection between individual decreases, $i.e$, $T_{\sigma}<T$ and its
minimal value $T_{\sigma}=\beta$ corresponds to the extreme
case of fully disconnection $\sigma=1$ and $t_r=t_b-1$. As a
consequence if $0<\beta<T_c$, the values of the parameters of our
strategy can be tuned to stop the epidemic spreading.

\begin{figure}[h]
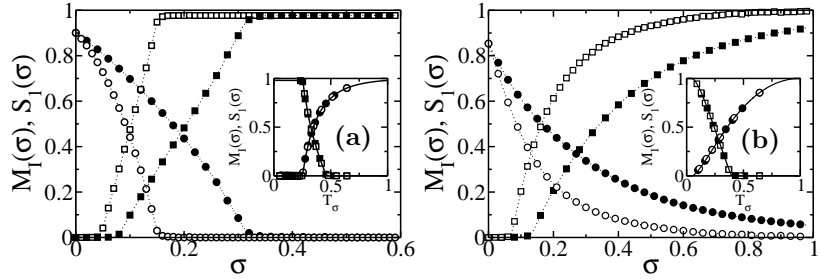

\centering
\vspace{1.0cm}
  \begin{overpic}[scale=0.20]{vbmbFig14.eps}
    \put(80,32){{\bf{(a)}}}
  \end{overpic}\vspace{0.5cm}
  \begin{overpic}[scale=0.20]{vbmbFig15.eps}
    \put(80,32){\bf{(b)}}
  \end{overpic}\vspace{0.5cm}
\caption{$M_{I}(\sigma,t_{b})\equiv M_{I}(\sigma)$ ($\circ$) and
  $S_{1}(\sigma,t_{b})\equiv S_{1}(\sigma)$ ($\square$) vs. $\sigma$
  for $N=10^{5}$, $t_r=20$ and $\beta=0.05$ in an ER network with $\langle
  k \rangle =4$ (a) and SF with $\lambda=2.63$, $k_{min}=2$ and
  $\langle k \rangle =4$ (b) for $t_b=10$ (empty symbols) and $t_b=19$
  (filled symbols). Dotted lines are given as guides for the eye. In
  the insets we show $M_{I}(\sigma,t_{b})$ and $S_{1}(\sigma)$ from
  the main plot as functions of $T_{\sigma}$ and the curves
  $M_{I}(\sigma)$ and $S_1(\sigma)$ obtained from percolation theory
  (solid lines), which show the mapping between the ISD strategy and
  percolation. Our simulations were averaged over $10^4$
  realizations.}\label{Ni_ER}
\end{figure}

In order to determine the effectiveness of the ISD strategy, we plot
the epidemic size $M_{I}(\sigma;t_{b})\equiv M_{I}(\sigma)$ and the
size of the functional susceptible network $S_{1}(\sigma;t_{b})\equiv
S_{1}(\sigma)$ as a function of $\sigma$ for ER and SF networks for
different values of $t_{b}$ and $t_r=20$. In Fig.~\ref{Ni_ER}, we can
see that $M_{I}(\sigma)$ decreases as $\sigma$ and $t_{b}$ increase
compared to the static case $M_{I}(0)$. For the SF network the
free-epidemic phase ($M_{I}(\sigma)=0$) is only reached for higher
values of $t_b$ and $\sigma$ than for ER networks. In any case, for
both homogeneous and heterogeneous networks, the strategy is
successful in protecting a giant susceptible component, for high
values of $\sigma$ and $t_b$.

Similarly to the disorder strategy, in this model $T_{\sigma}$ maps
with a percolation process (see the insets of Fig.~\ref{Ni_ER}), and
also when $T_{\sigma}=T^{*}$, the size distribution of the susceptible
clusters behaves as $n_{s}\sim s^{-2.5}$ (not shown here). In turn, in
the ISD strategy the susceptible individuals change dynamically their
connectivities with the infected neighbors, reducing the contact time
between them. This generates an adaptive topology~\cite{Gro_01} in
which the susceptible ones aggregate into clusters that produce a
resistance to the disease. Therefore in the ISD strategy there is also
a ``susceptible herd behavior''.

\begin{figure}[ht]
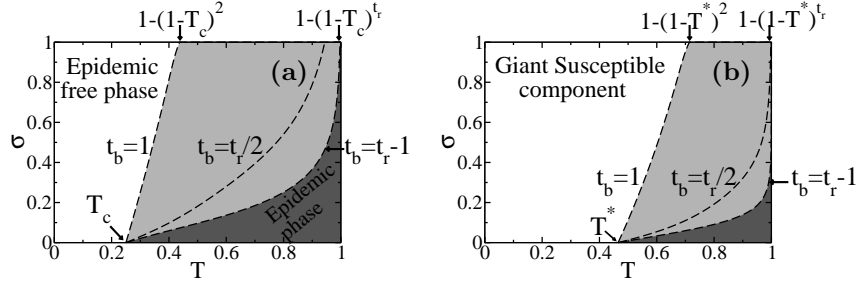

\centering
 \begin{overpic}[scale=0.20]{vbmbFig16.eps}
    \put(65,50){{\bf{(a)}}}
  \end{overpic}\hspace{0.3cm}
  \begin{overpic}[scale=0.20]{vbmbFig17.eps}
    \put(65,48){\bf{(b)}}
  \end{overpic}\vspace{0.5cm}
\caption{Plot of the epidemics phase (a) and GSC phase (b) in the
  plane $\sigma-T$ for $t_r=20$ and static $T_c=0.25$, where $T$
  corresponds to the transmissibility in a non adaptive network . The
  dashed lines correspond to the critical threshold transmissibility
  $T_{\sigma}=T_{c}$ (a) and $T_{\sigma}=T^{*}$ (b) for (from left to
  right) $t_b=1$, $t_{b}=t_{r}/2$ and $t_b=t_r-1$. For $t_{b}=1$ and
  $\sigma=1$, $T=1-(1-T_{c})^{t_r/([(t_r+1)/2]+1)}$ or $T\approx
  1-(1-T_{c})^{2}$ is the maximum intrinsic transmissibility for which
  the epidemic phase disappears when the ISD strategy is
  applied.}\label{Fase}
\end{figure}

In order to study the performance of the strategy protecting a GSC or
preventing an epidemic phase, in Fig.~\ref{Fase} we plot the plane
$\sigma-T$ [where $T\equiv T(\sigma=0)$] for different values of $t_b$,
using Eq.~(\ref{Ec.Trans}) for $T_{\sigma}=T_c$ and
$T_{\sigma}=T^{*}$.
\noindent

In Fig~\ref{Fase} (a-b) starting from the case without strategy (line
$\sigma=0$) the epidemic phase and the phase without GSC shrink when
$\sigma$ and $t_b$ increase. Note that the light-gray area, delimited
between the curves which corresponds to the extreme blocking periods $t_{b}=1$ and
$t_{b}=t_{r}-1$, displays the region of parameters controlled by the
intervention strategy. In particular, given $t_b$ and $t_r$, the
maximum intrinsic transmissibility at which the strategy can prevent
an epidemic phase or protect a GSC can be obtained using
Eq.~(\ref{sigma1}) for $T_{\sigma}=T_{c}$ or $T_{\sigma}=T^{*}$
respectively, and $\beta=1-(1-T)^{1/t_r}$. On the other hand, note
that in pure SF networks with $2<\lambda\leq 3$ and $k_{max}=\infty$,
$T_c=0$, which implies that the strategy cannot halt the epidemic
spreading for any value of the intrinsic transmissibility. However,
$T^{*}$ is still finite on these topologies. Therefore, the ISD
strategy can always protect the functional network for diseases with
$T<1-(1-T^{*})^{t_r}$.

For the disorder strategy, we can reach similar conclusions because
it is expected that the magnitudes in the steady state will behave in
the same way for any strategy that is governed by the
transmissibility. However, as we will show below, the evolution
towards the steady state is different in both strategies.

\section{Comparison between the ISD and the quenched disorder strategy}
\begin{figure}[ht]
 \centering
\vspace{0.5cm}
\includegraphics[scale=0.23]{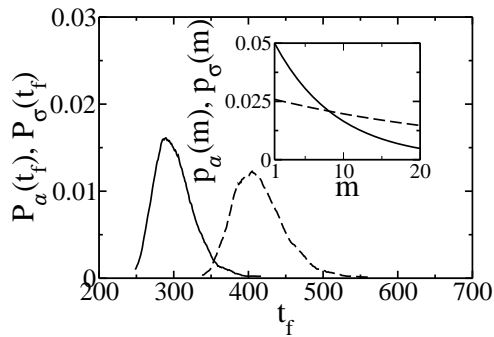}
 \caption{Distribution of final times $t_{f}$ in an epidemic
   spreading, with $N=10^5$, $\beta=0.05$ and $t_r=20$ in a ER network
   with $\langle k \rangle =4$ for the quenched disorder strategy with
   $a=1.5$ (dashed line) and ISD strategy (solid line) with $t_b=19$
   and $\sigma=0.0695$. Both strategies have the same effective
   transmissibility value $T_a=T_{\sigma}\approx 0.39$. The final
   average time for the quenched disorder strategy is $\langle t_{f}
   \rangle=406$ and $\langle t_f \rangle=290$ for the ISD strategy,
   giving a ratio between these times of $1.38$. In the inset, we show
   the probability that an active link transmits the disease at time
   $m$, since it appears (with $1\leq m \leq t_r$). The average time
   to traverse the disease is $\langle m \rangle=3.75$ for the
   quenched disorder strategy and $\langle m \rangle=2.67$ for the ISD
   strategy, and the ratio is $1.40$ that is compatible with the ratio
   between the most probable final time for both strategies. }
 \label{DistribFin}
\end{figure}

In Fig.~\ref{DistribFin} we plot the distribution of the duration time
$t_f$ of an epidemic for the ISD strategy $P_{\sigma}(t_{f})$ and the
quenched disorder strategy $P_{a}(t_{f})$ for the same value of
transmissibility $T_a=T_{\sigma}$.

From the figure, we can see that the quenched disorder strategy
generates larger duration times of the epidemic, $i.e.$, the disease
spreading is slower than in the ISD strategy, which shows that the
transmissibility does not govern magnitudes involved in the
dynamical behavior. However, the discrepancy between the strategies
can be explained from the transmissibility's terms of
Eqs.~(\ref{Tdisorder}) and (\ref{Ec.Trans}).

Lets denote the first time step of the existence of an active link as
$m=1$. Then using Eq.~(\ref{Tdisorder}), the probability $p_{a}(m)$
that the infected individual transmits the disease at time step $1\leq
m\leq t_r$, for the disorder quenched strategy, is given by
\begin{eqnarray}
p_{a}(m)&=& \int_{e^{-a}}^{1} \frac{\beta w (1-\beta w)^{m}}{a w}dw\nonumber\\
&=&\frac{\left(1-\beta e^{-a}\right)^{1+m}-\left(1-\beta\right)^{1+m}}{a\left(1+m\right)}.
\end{eqnarray}
Similarly, for the ISD strategy, the probability $p_{\sigma}(m)$ that
the infected individual transmits at time $1\leq m\leq t_r$ is,
\begin{eqnarray}
p_{\sigma}(m)&=&\beta(1-\beta)^{m-1}(1-\omega)^{m-1}+\beta \sum_{u=1}^{\left[\frac{m-1}{t_{b}+1}\right]}\binom{m-u\;t_{b}-1}{u}\sigma^{u}\times\nonumber\\
&&(1-\sigma)^{m-1-u(t_{b}+1)}(1-\beta)^{m-1-u\;t_{b}},
\end{eqnarray}
From these probabilities, we compute the average time steps $\langle m
\rangle$ that takes to the disease to traverse an active link for
several values of the parameters from both strategies, and we obtain
that in the quenched disorder strategy the disease needs more time to
infect a susceptible individual than in the ISD strategy (see the
inset in Fig.~\ref{DistribFin}). Thus it is expected that the final
times $t_{f}$ in the former will be longer than in the latter. On the
other hand, the ratio between the average times $\langle m \rangle$ is
compatible with ratio between the most probable final times of the
distributions $P_a(t_{f})$ and $P_{\sigma}(t_{f})$. These results show
that we can use minimal information, specifically the terms of the
transmissibility in order to determine if the strategy slows down the
epidemic spreading. Since one of the goals of the health authorities
is to have more time to intervene, the average time $\langle m
\rangle$ could be used to compare, design or optimize mitigation
strategies.

\section{Summary}
Percolation theory offers the possibility to explain the epidemic
spreading and mitigation strategies in geometrical terms. In this brief
review, we focused on the applications of percolation theory for the
studying of social distancing strategies against the epidemic
spreading of the SIR model.

We described the dilution of the network composed by susceptible
individuals due to the disease spreading as a ``node void
percolation'' process, and remark its importance in the development of
strategies that aims to protect the functional network.

Using the SIR model for the disease propagation, we presented two
social distancing strategies: the quenched disorder strategy and the
intermittent social distancing strategy. We found that both strategies
can control the effective transmissibility in order to protect the
society. In particular, we described the protection of the GSC through
the formation of a susceptible herd behavior. On the other hand, we
showed that while the effective transmissibility control the final
fraction of recovered individuals and the size of the GSC, it does not
control observables that depends on the dynamical evolution of the
process, such as the distribution of the duration of an epidemic.

One of the advantages of having two strategies that map with
percolation theory is that we can fix the transmissibility in order to
compare them and highlight the features of each strategy. Thus, for
example, the knowledge of the mean time $\langle m \rangle$ that a
disease requires to traverse an active link, can be used to determine
which strategy is better in delaying the epidemic spreading. Using the
terms of the transmissibility, we showed that the quenched disorder
strategy increases this average time, and thus the epidemic spreading
is delayed compared to the ISD strategy. Our results show that a
disorder strategy has a deeper effect on the spreading dynamics than a
local adaptive topology.

Our findings could themselves have important applications for
improving or designing mitigation strategies, since new strains of
bacterias and viruses are continuously emerging or reemerging in
multi-drug resistant forms, demanding the development of
non-pharmaceutical intervention.

\section*{Acknowledgments}
This work was financially supported by UNMdP and FONCyT (Pict
0293/2008). We thank C. E. La Rocca for useful comments and
discussions.

\appendix{Edge-Based Compartmental Model}

The edge-based compartmental model~\cite{Mil_04,Mil_03,Val_02}, is a
new theoretical framework to describe the dynamic of the disease
spreading in the SIR model. Using this approach we can obtain the
relation between $V^{s}$ and $S_{1}$.

For clarity, we return to the SIR terminology, in which a void node
corresponds to a susceptible individual and the node belonging to the
giant percolating cluster (in a branching process) corresponds to a
recovered individual.

In order to compute $S_1$, we first calculate the fraction of
susceptible individuals and then subtract the fraction of susceptible
individuals belonging to finite size clusters.

Consider an epidemic disease in the steady state. We randomly choose a
link and then give a direction to that link, in which the node in the
target of the arrow is called the root, and the base is its
neighbor. Denote $\theta$ as the probability that the neighbor has
never transmitted the disease to the root, due to the fact that the
neighbor is: (i) susceptible, or (ii) recovered, but he has never
transmitted the disease to the root during its infectious period,
$i.e.$
\begin{eqnarray}
\theta=V^{s}+(1-p)f_{\infty}(p).
\end{eqnarray}
where $p=T$.
Therefore the probability that the root with connectivity $k$ is
susceptible is $\theta^k$, $i.e$, an individual is susceptible only if
none of his neighbors have transmitted the disease to him. Then,
considering all the connectivities $k$, the fraction of susceptible
individuals in the steady state is $G_{0}(\theta)$. Note that $V^s$
can also be related to $\theta$, since reaching a node through a link,
it is susceptible only if none of its outgoing neighbors are connected
to the giant recovered cluster, that is,
\begin{eqnarray}\label{eqs111}
V^{s}=G_1(\theta).
\end{eqnarray}
On the other hand, if we define $\omega$ as the probability that the
neighbor is (i) susceptible but it does not belong to a GSC, or (ii)
recovered, but he has never transmitted the disease to the root during
its infectious period, then we have,
\begin{eqnarray}\label{eqs111a}
\omega=G_{1}(\omega)+(1-p)f_{\infty}(p),
\end{eqnarray}
where $G_{1}(\omega)$ is similar to $V^s$, but restricted only to
susceptible neighbors who belong to finite susceptible size
clusters (see Eq.~\ref{eqs111}).

\begin{figure}[ht]
 \centering
\vspace{0.5cm}
\includegraphics[scale=0.23]{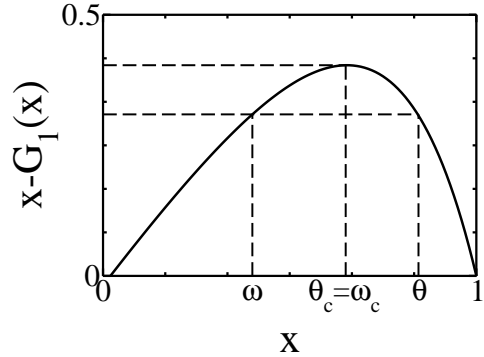}
 \caption{Schematic of the behavior of Eq.~(\ref{eqs112}). For
   $\theta\neq \omega$ we have two solutions. When $\theta$ reaches
   the maximum of the function $x-G_{1}(x)$, $\theta_{c}=\omega_{c}$,
   the giant susceptible component is destroyed (see
   Eq.~\ref{eqs113}). The dashed lines are used as a guide to show the
   possible solutions of Eq.~(\ref{eqs112}). }
 \label{figtheOme}
\end{figure}

Then, from Eqs.~(\ref{eqs111}) and (\ref{eqs111a}) we obtain
\begin{eqnarray}\label{eqs112}
\theta - G_{1}(\theta)&=&\omega -G_{1}(\omega).
\end{eqnarray}
Note that both hand sides of Eq.~(\ref{eqs112}) has the form
$x-G_1(x)$. In Fig.~\ref{figtheOme}, we illustrate the solution of this
equation.

Finally, for a given value of $V^{s}$, we can solve
Eqs.~(\ref{eqs111})~and~(\ref{eqs112}), in order to compute the
relative size of the GSC, as
\begin{eqnarray}
S_{1}&=&G_{0}(\theta)-G_{0}(\omega).\label{eqs113}
\end{eqnarray}
where $G_{0}(\omega)$ is the fraction of void
nodes belonging to finite void clusters (see Ref.~\cite{Val_02} for
details).

On the other hand, from Eq.~(\ref{eqs112}) we can obtain the critical
value $V^{s}_{c}$ at which $S_1$ vanishes, $i.e.$, when
$G_{0}(\theta)=G_{0}(\omega)$. Note that this happens only when
$\theta=\omega$, because $G_{0}(x)$ is an strictly increasing
function. In addition, since $\theta$ and $\omega$ fulfills
Eq.~(\ref{eqs112}), $\theta=\omega$ only at the maximum of
$x-G_{1}(x)$ (see Fig.~\ref{figtheOme}). Then, denoting the maximum as
$\theta_{c}=\omega_c$, we have that
\begin{eqnarray}\label{Eqmax1}
\left[x-G_{1}(x)\right]^{'}\big|_{\theta_c}=0,
\end{eqnarray}
then,
\begin{eqnarray}
\theta_{c}=\left(G_{1}^{'}\right)^{-1}(1).
\end{eqnarray}
Thus using Eq.~(\ref{eqs111}), the critical threshold of the
susceptible network is
$V^{s}_{c}=G_{1}(\theta_c)=G_{1}\left[\left(G_{1}^{'}\right)^{-1}(1)\right]$,
which for ER networks $V^{s}_{c}=1/\langle k \rangle$.

Finally, we show the mean field exponent of $S_1$ as a function of $V^s$.

Near the critical threshold of the susceptible network, the values of
$\theta$ and $\omega$ from Eq.~(\ref{eqs112}) are near to $\theta_c$,
in which we can approximate the function $x-G_{1}(x)$ as a
parabola. Thus $x-G_{1}(x)\approx a-b/2(x-\theta_c)^2$, where $a$ and
$b$ are constants. Doing some algebra on Eq.~(\ref{eqs112}) around
$\theta_c$, we obtain
\begin{eqnarray}
 \left | \omega - \theta_{c} \right | \approx \left | \theta -
 \theta_{c} \right |,
\end{eqnarray}
$i.e.$, $\theta_c$ is in the middle between $\omega$ and
$\theta$. Rewriting $\theta$ and $\omega$ as $\omega\approx
\theta_{c}-\Delta$ and $\theta \approx \theta_{c}+\Delta$, with
$\Delta \ll 1 $, then near criticality, Eq.~(\ref{eqs113}) can be
approximated by
\begin{eqnarray}\label{plaw1}
S_{1}&\approx&G_{0}(\theta_{c}+\Delta)-G_{0}(\theta_{c}-\Delta)\nonumber\\
&\approx&2G_{0}^{'}(\theta_c)(\theta-\theta_{c}).
\end{eqnarray}
On the other hand, near criticality we have that
\begin{eqnarray}\label{plaw2}
V^{s}-V^{s}_{c}&=&G_1(\theta)-G_{1}(\theta_c)\nonumber\\
&\approx&G_1(\theta_c+\Delta)-G_{1}(\theta_c)\nonumber\\
&\approx& G_{1}^{'}(\theta_c)(\theta-\theta_c).
\end{eqnarray}
Therefore, using the relations~(\ref{plaw1})~and~(\ref{plaw2}), we obtain
\begin{eqnarray}
S_1\sim (V^{s}-V^{s}_{c})^{\beta},
\end{eqnarray}
with $\beta=1$, that is a MF exponent. Note that we have not made any
assumption on the form of $G_{1}(x)$ or $G_{0}(x)$. Thus, this result
is valid for homogeneous and heterogeneous networks.

\end{document}